\documentclass[prb,
12pt,
superscriptaddress,showpacs,amsmath,amssymb]{revtex4}
\usepackage{amsfonts}
\usepackage{bm}
\usepackage{verbatim}

\usepackage{graphicx}

\begin{document}

\author{G.E.~Volovik}
\affiliation{Low Temperature Laboratory, Aalto University,  P.O. Box 15100, FI-00076 Aalto, Finland}
\affiliation{Landau Institute for Theoretical Physics, acad. Semyonov av., 1a, 142432,
Chernogolovka, Russia}

\title{Dimensionless physics: continuation}

\date{\today}

\begin{abstract}
Several approaches to quantum gravity (including the model of superplastic vacuum; Diakonov tetrads emerging  as the bilinear combinations  of the fermionis fields; $BF$-theories of gravity; and effective acoustic metric)  suggest that in general relativity  the metric must have dimension 2, i.e. $[g_{\mu\nu}]=1/[L]^2$, irrespective of the dimension of spacetime. This leads to the "dimensionless physics" discussed in the review paper \cite{Volovik2021}.  Here we continue to exploit this unusual dimension of the metric.

\end{abstract}
%\pacs{
%}

\maketitle

\tableofcontents

\section{Introduction: elasticity tetrads, quantum mechanics and general relativity}

Here we discuss some issues related to the dimensions of physical variables, which follow from the peculiar dimensions of the tetrads and metric.  The particular example is represented by the elasticity tetrads -- the  translational gauge fields, which have dimension $1/[L]$,  The corresponding metric field, which is the bilinear combination of the tetrad fields,  has dimension  $1/[L]^2$. There are the other approaches to gravity, which also result in the same  dimensions. All this leads to the "dimensionless physics".\cite{Volovik2021} 
In principle, any given physical theory can be rewritten in the dimensionless form. This is trivial from the mathematical point of view, since one can always find the proper system of units. However, with the dimensional tetrads, the physics automatically becomes dimensionless. Here we continue to exploit the dimensionful tetrads. In particular we consider how such tetrads make connection between quantum mechanics and general relativity.

\section{Dimensions of tetrads and metric}

\subsection{Dimensional tetrads}

Let us start with the model of the superplastic vacuum.
The 3 + 1-dimensional vacuum crystal is the plastic (malleable) medium,\cite{KlinkhamerVolovik2019} which is
 described in terms of the so-called elasticity tetrads:\cite{Dzyaloshinskii1980,NissinenVolovik2019,Nissinen2020,Burkov2021} 
\begin{equation}
E^a_\mu= \frac{\partial X^a}{\partial x^\mu}\,,
\label{ElasticityTetrads}
\end{equation}
where equations $X^a(x)=2\pi n_a$ are equations of the (deformed) crystal planes. Since  the functions $X^a$ play the role of the geometric $U(1)$ phases and thus are dimensionless, the elasticity tetrads play the role of the gauge fields (translation gauge fields). That is why tetrads have the same dimension 1 as the dimension of gauge fields:
\begin{equation}
[E^a_\mu] = \frac{1}{[L]}\,.
\label{ElasticityTetradsDim}
\end{equation}
The dimension $n$ of quantity $A$ means  $[A]=[L]^{-n}$, where $[L]$ is dimension of length.   Tetrads with dimension 1 lead to dimensionless physics,\cite{Volovik2021} 
which we discuss here. 

Note that the $U(1)$ gauge fields, which form the dimensional tetrads, are not necessarily related to translations and the crystalline structure of the vacuum. The latter is only one of the possible sources of the dimensional tetrads. In particular, the dimension 1 tetrads appear in the Diakonov theory,\cite{Diakonov2011,VladimirovDiakonov2012,VladimirovDiakonov2014,ObukhovHehl2012}
see Sec. \ref{WeylDirac}. 

Note also, that in principle, the matrix $E^a_\mu$ is not necessarily quadratic. The extension of tetrads to the rectangular vilebein is considered in Ref.\cite{Volovik2022b}.

\subsection{Dimensional metric}

Elasticity tetrads in Eq.(\ref{ElasticityTetrads}) give rise to the metric with lower indices, which is the following bilinear combination of tetrads:
\begin{equation}
g_{\mu\nu}=\eta_{ab}E^a_\mu E^b_\nu \,.
\label{MetricElasticity}
\end{equation}
This covariant metric tensor has dimension $n=2$, while the contravariant metric $g^{\mu\nu}$ has dimension $n=-2$:
 \begin{equation}
[g_{\mu\nu}] =\frac{1}{[L]^2}\,\,,\,\, [g^{\mu\nu}] =[L]^2\,.
\label{MetricDimension}
\end{equation}
Another example of the metric with the dimension 2 is the effective metric describing the propagation of Goldstone modes (phonons) in moving superfluids,  the so-called acoustic metric,\cite{Unruh1981,Visser1998}  see  Sec. \ref{SecAcoustic}.

The tetrad determinant has dimension 4 in the 4-dimensional spacetime:
 \begin{equation}
[e]=[\sqrt{-g}\,] =\frac{1}{[L]^4} \,.
\label{DimensionDeterminant}
\end{equation}
In the $N$-dimensional spacetime, the dimensions of the metric elements are the same as in Eq.(\ref{MetricDimension}), while the tetrad  determinant has dimension $N$:
 \begin{equation}
[e]=[\sqrt{-g}\,] =\frac{1}{[L]^N} \,.
\label{DimensionDeterminantN}
\end{equation}

\subsection{Action and Lagrangian}

Eq.(\ref{DimensionDeterminantN}) makes the spacetime integration dimensionless:
\begin{equation}
\big[ \int d^N x \sqrt{-g}\,\big]=[1]=0 \,.
\label{DimensionlessIntegral}
\end{equation}
Since the action is dimensionless, Eq.(\ref{DimensionlessIntegral}) leads to the dimensionless Lagrangian, $[{\cal L}]=[1]=0$:
\begin{equation}
\big[S\big]=\big[ \int d^N x \sqrt{-g}\,{\cal L}\big]=\big[ \int d^N x \sqrt{-g}\,\big]\cdot \big[{\cal L}\big]=[1]\cdot [1]=[1] \,.
\label{DimensionlessLagrangian}
\end{equation}

\subsection{Dynamics of classical point particle}

Eq.(\ref{MetricElasticity}) gives rise to the dimensionless interval:
 \begin{equation}
ds^2=g_{\mu\nu}dx^\mu dx^\nu \,\,, \,\,  [s^2]= \frac{1}{[L]^2} \cdot [L]^2=[1] =0\,.
\label{DimensionInterval}
\end{equation}
The interval describes the classical dynamics of a point particle with action:
 \begin{equation}
S=M\int ds \,.
\label{particleAction}
\end{equation}
The variation of action leads to the Hamilton--Jacobi equation expressed in terms of the contravariant metric $g^{\mu\nu}$:
 \begin{equation}
g^{\mu\nu}\partial_\mu S \partial_\nu S + M^2 =0\,.
\label{HJ}
\end{equation}
Since the action is dimensionless, the dimensionless interval makes the particle mass  $M$ in  Eq.(\ref{particleAction}) also dimensionless, $[M]=[1]=0$. Then both terms in Eq.(\ref{HJ}) are dimensionless. This is valid for the arbitrary dimension $N$ of spacetime.

In the spacetime crystal, the interval between the events is counted in terms of the lattice points, and this is the geometric reason why the interval is dimensionless. In case of the other possible origins of the dimensionless interval, this demonstrates that the interval determines the dynamics of the particle, rather than the geometric distance.
In the first case the dynamics comes from geometry, while in the second case the geometry follows from dynamics.
That is why there is the difference between the dimension $[L]$ of the geometric quantity, the length, and the zero dimension $[L]^0$ of the dynamical quantity, the interval.

\section{Dimensions of quantum fields in curved space}

\subsection{Universal dimension of scalar fields}

Let us consider the quadratic terms in the action for the scalar field $\Phi$ in the $N$-dimensional spacetime:
\begin{equation}
S=\int d^N x\,\sqrt{-g} \,\left(  g^{\mu\nu} \nabla_\mu \Phi^*  \nabla_\nu \Phi +M^2|\Phi|^2 \right)\,,
\label{scalar}
\end{equation}
Taking into account equations (\ref{MetricDimension}) and (\ref{DimensionlessIntegral}), one obtains that the scalar field is dimensionless, $[\Phi]=[1]=0$, for arbitrary spacetime dimension $N$. So, it has the same zero dimension as the Goldstone modes, see also Sec. \ref{SecAcoustic}. This universal zero dimension differs from the $N$-dependent dimension of scalar fields in the conventional approach, where the dimension is   $n=(N-2)/2$.

\subsection{Dimension of wave function in quantum mechanics}

 Expanding the Klein-Gordon equation for scalar $\Phi$ in Eq.(\ref{scalar}) over $1/M$ one obtains the non-relativistic Schr\"odinger action in curved space-time. Let us for simplicity consider the space-time independent  metric and assume that $g^{0i}=0$. Then introducing the Schr\"odinger wave function $\psi$:
\begin{equation}
\Phi({\bf r},t) = \frac{1}{\sqrt{M}}\exp\left(i Mt /\sqrt{-g^{00}}\right)\psi({\bf r},t)  \,,
\label{eq:PhiPsi}
\end{equation}
 one obtains the Schr\"odinger-type  action in the form
\begin{eqnarray}
S_\text{Schr}=\int d^3x dt  \sqrt{-g}\, {\cal L} \,,
\label{eq:SchroedingerAction}
\\
2{\cal L}= 
i\sqrt{-g^{00}} \left(\psi \partial_t \psi^*-\psi^* \partial_t \psi\right) 
%  \frac{ig^{0k}}{\sqrt{-g^{00}} }\left(\Psi \nabla_k \psi^*-\psi^* \nabla_k \Psi\right)
+\frac{g^{ik}}{M}\nabla_i\psi^* \nabla_k \psi   \,.
\label{eq:SchroedingerEq}
\end{eqnarray}
The normalization condition  for the wave function $\psi$ is:
\begin{equation}
\int d^3 r  \,\sqrt{\gamma}\, |\psi |^2 = 1\,,
\label{eq:PhiPsi}
\end{equation}
where  $\sqrt{\gamma}= \sqrt{-g} \sqrt{-g^{00}}$ is the determinant of the space part of the metric. This corresponds to the particle number conservation in the nonrelativistic quantum mechanics, see e.g. Eq.(13) in Ref. \cite{Kabel2022}.

Since the dimension of this determinant is $[\sqrt{\gamma}\,]=\frac{1}{[L]^3}$, the wave function $\psi$ is dimensionless. The dimension $[\psi]=[1]$ can be also obtained directly from Eqs. (\ref{eq:SchroedingerAction}) and (\ref{eq:SchroedingerEq}), since the Lagrangian is dimensionless, $[{\cal L}]=[1]$.  This is distinct from the conventional Schr\"odinger equation without gravity, where the dimension of the wave function is $[\psi]=[L]^{-3/2}$ (or $[\psi]=[L]^{-(N-1)/2}$ for the $N$ dimensional spacetime). Inclusion of gravity in the form of the vacuum crystal tetrads or in terms of the Diakonov tetrads provides the natural zero dimension for the  probability amplitude in quantum mechanics, $[\psi]=[1]$, for any spacetime dimension $N$. 

The same result can be obtained using the Dirac  bra–ket notations. The overlap of the quantum states is naturally dimensionless:
\begin{equation}
<{\bf r}\,|\,{\bf r}'> = \frac{1}{\sqrt{\gamma}}\,\delta({\bf r}-{\bf r}') \,.
\label{BraKet1}
\end{equation}
Then for the wave function
\begin{equation}
\psi({\bf r})= <{\bf r}\,|\, \psi>  \,\,, \,\, |\, \psi> =
\int d^{N-1} r  \,\sqrt{\gamma} \,\psi({\bf r}) |\,{\bf r}>
 \,,
\label{BraKet2}
\end{equation}
one obtains Eq.(\ref{eq:PhiPsi}) for normalization:
\begin{equation}
1=<\psi\,|\, \psi\,> =\int d^{N-1} r  \,\sqrt{\gamma}\, |\psi |^2 \,.
\label{eq:PhiPsiBraKet}
\end{equation}
This is valid only if the wave function is dimensionless, $[\psi]=[1]=0$. This is the consequence of the presence of the  metric field in Eq.(\ref{eq:PhiPsiBraKet}), which demonstrates the close connection between quantum mechanics and general relativity.

Note, that the action (\ref{eq:SchroedingerAction}) and Lagrangian (\ref{eq:SchroedingerEq}) do not contain $\hbar$.
The role of $\hbar$ in the conventional relation between the energy levels and frequency, $E_m -E_n=\hbar \omega_{mn}$, is now played by $\sqrt{g^{00}}$ in the red shift equation,\cite{Volovik2009} $M_m - M_n=\sqrt{g^{00}}\, \omega_{mn}$. The dimenional metric leads to the difference between  the dimensional frequency, $[\omega_{mn}]=1/[L]$, and the  dimensionless mass:
\begin{equation}
 [M]=[\sqrt{g^{00}}] [\omega] =[L] \cdot  \frac{1}{ [L]} =[1]=0 \,.
\label{FrequencyDimension}
\end{equation}
 
\subsection{Weyl and Dirac fermions}
\label{WeylDirac}

On the other hand, the natural dimension of tetrads, $[E^a_\mu]=1/[L]$, can be obtained without consideration of the model of superplastic vacuum or Diakonov theory. It is enough to use the natural zero dimension of wave functions, $[\psi]=[1]$, which gives rise to the dimensionless Weyl and Dirac fields, $[\Psi]=[1]=0$. For example, let us consider the action for massless Dirac fermions
\begin{equation}
S=\int d^4x\,  e\, e^\mu_a \bar\Psi  \gamma^a \nabla_\mu \Psi \,,
\label{Fermions}
\end{equation}
where $e$ is the tetrad determinant. Since the action is dimensionless, then assuming that the quantum field operators $\Psi$ are dimensionless,   $[\Psi]=[1]=0$,  one obtains $ [e\,e^\mu_a] =1/[L]^3$, which gives the dimensional tetrads:
\begin{equation}
 [e^\mu_a] = [L]\,\,, \, [E_\mu^a] =\frac{1}{ [L]}  \,\,, \,\ [e] = \frac{1}{[L]^4}\,.
\label{FermionDimension}
\end{equation}

The Hamiltonian  for massless Dirac fermions has dimension 1, i.e.  the same as the dimension of frequency:
\begin{equation}
H=\int_{x_0={\rm const}} d^3 r\,  e\, e^i_a \bar\Psi  \gamma^a \nabla_i \Psi  \,\,, \,\ [H] =[\omega]= \frac{1}{[L]}\,.
\label{FermionHamiltonian}
\end{equation}
The dimension of the Hamiltonian does not coincide with the dimension of mass $M$, which is dimensionless, see Eq.(\ref{FrequencyDimension}).

Although the fermionic fields are dimensionless, the current density operator of these fields has the conventional dimension 3. This is due to the dimension of tetrads:
\begin{equation}
j^\mu =e e^\mu_a \bar\Psi  \gamma^a   \Psi \,\,, \,\, [j^\mu]=[e] [e^\mu_a] = \frac{1}{[L]^3}\,.
\label{current}
\end{equation}

The dimensionful tetrads also appear in the Diakonov theory,\cite{Diakonov2011,VladimirovDiakonov2012,VladimirovDiakonov2014,ObukhovHehl2012}
where tetrads emerge as the bilinear combinations of the fermionic fields:
\begin{equation}
E_\mu^a \, \propto \, <\bar\Psi  \gamma^a \nabla_\mu \Psi >\,\,, \, [E_\mu^a] =\frac{1}{ [L]} \,.
\label{bilinear}
\end{equation}
In this approach the metric $g_{\mu\nu}$ is the quadrilinear combination of the fermionic fields, $<\bar\Psi  \Psi \bar\Psi   \Psi >$.
This approach also allows to introduce the rectangular vilebein discussed in Ref.\cite{Volovik2022b}, since the spin $a$ and coordinate $\mu$ spaces may have different dimensions.

\subsection{Gauge fields}
\label{SecGauge}

The action for the $U(1)$ gauge field in the $N$-dimensional spacetime is:
\begin{equation}
S \sim \int d^Nx\,\sqrt{-g}\, g^{\mu \nu} g^{\alpha \beta} F_{\mu\alpha}F_{\nu\beta} \,.
\label{gauge}
\end{equation}
In case of the conventional dimensionless tetrads, the action in Eq.(\ref{gauge}) is dimensionless only for $N=4$. 
The dimensionless prefactor $1/\alpha$ in Eq.(\ref{gauge}) for $N=4$ -- the electric permittivity of the quantum vacuum -- will be discussed in Sec. \ref{ADMsec}.

With dimensionful tetrads the action (\ref{gauge}) is dimensionless for arbitrary $N$, since
\begin{equation}
 [g^{\mu \nu}]=[L]^2\,\, \,, \,\, [F_{\mu\nu}] =\frac{1}{ [L]^2}  \,\,, \,\ [\sqrt{-g} ] = \frac{1}{[L]^{N}}\,.
\label{GaugeDimension}
\end{equation}
The same is valid for the non-Abelian gauge fields.

\subsection{Dimension of acoustic metric}
\label{SecAcoustic}

The effective acoustic metric describes propagation of sound in a non-homogeneous flowing fluid,\cite{Unruh1981,Visser1998} 
and also phonons in moving superfluids and other Goldstone modes, such as magnons and collective modes of magnon Bose condensate.\cite{NissinenVolovik2017} The action for the propagating Goldstone mode (the phase $\phi$ of the Bose condensate) is similar to the action (\ref{scalar}) for the massless scalar field:
\begin{equation}
S=\int d^4 x\,\sqrt{-\tilde g} \, \tilde g^{\mu\nu} \nabla_\mu \phi  \nabla_\nu \phi \,.
\label{Acoust}
\end{equation}
There some differences between Eq.(\ref{Acoust}) and Eq.(\ref{scalar}):
(i) the phase $\phi$ of the superfluid order parameter is dimensionless in any approach; (ii) $\tilde g^{\mu\nu}$ is the emergent (effective) acoustic metric, which is determined by the velocity and density of the superfluid liquid; (iii) 
the real gravitational field is not taken into account and thus the conventional dimensions can be used.

From the action (\ref{Acoust}) it follows that the effective contravariant metric
$\tilde g^{\mu\nu}$ has the conventional dimension $-2$, i.e. $[\tilde g^{\mu\nu}]=[l]^2$. This can be also seen form the dependence of  the effective interval on the hydrodynamic variables:\cite{Visser1998,Volovik2003}
\begin{equation}
d\tilde s^2=\tilde g_{\mu\nu}dx^\mu dx^\nu=\frac{n}{ms}[-s^2 dt^2 + (dx^i -v^idt)\delta_{ij}  (dx^j -v^jdt)]\,.
\label{AcousticInterval}
\end{equation}
Here $n$ is the density of atoms in the liquid; $m$ is the mass of the atom; $s$ is the speed of sound; and $v^i$ is the velocity of the superfluid liquid, which coincides with the shift vector $N^i$ in the Arnowitt-Deser-Misner formalism in Sec. \ref{ADMsec}.

Using the conventional dimensions of hydrodynamic quantities in Eq.(\ref{AcousticInterval}) one obtains the dimension 2 for the covariant metric $\tilde g_{\mu\nu}$:
\begin{equation}
[\tilde g_{\mu\nu}]=[n] \cdot \frac{1}{[m]}= \frac{1}{[l]^3} \cdot [l]= \frac{1}{[l]^2}  \,,
\label{AcousticDimension}
\end{equation}
and the dimensionless interval.
Such dimension of the effective metric follows solely from the dynamics of the superfluid. This provides another example when the geometry comes from dynamics.

The effective metric in Eq.(\ref{AcousticInterval}) depends on two "Planck" energy scales. One of them is related to the mass $m$ of the atom of the liquid and another one is determined by the average distance $a$ between the atoms in the liquid. Together they produce the effective Planck energy scale, which determines the vacuum energy (see Eq.(3.26) in Ref.\cite{Volovik2003}):
\begin{equation}
E_{\rm Planck}=E^{-1/2}_{\rm Planck\,1}E^{3/2}_{\rm Planck\,2}\,\,,\,\, E_{\rm Planck\,1}=ms^2\,\,,\,\,E_{\rm Planck\,2} = n^{1/3}=\frac{1}{a}\,.
\label{PlanckScales}
\end{equation}
In general relativity, these two Planck scales are considered to be equal: $E_{\rm Planck\,1}=E_{\rm Planck\,2}=E_{\rm Planck}$, where $E_{\rm Planck\,1}$ corresponds to the Planck mass, and $E_{\rm Planck\,2}$ corresponds to the inverse Planck length.

 \section{General relativity}
 
\subsection{Einstein action}

Let us consider the GR action on example of $q$-theory. This is the class of theories which avoid the cosmological constant problem: the huge contributions of zero point energy to the cosmological constant is cancelled in the equilibrium state of the vacuum due to thermodynamics.\cite{KlinkhamerVolovik2008a,KlinkhamerVolovik2008,KlinkhamerVolovik2022} 
For the particular $q$-theory on the four-dimensional (4D) ``brane'' the action is:\cite{KlinkhamerVolovik2016,Klinkhamer2022}
\begin{eqnarray}
\hspace*{-10mm}
S &=&- \int
d^4x\,\sqrt{-g}\,\left[\epsilon(q)+\frac{R}{16\pi G_{N}(q)} +\Lambda_0+\mathcal{L}^{M}[\psi,q]\right]
\nonumber\\[2mm]
&&+ \mu \int\,d^4x \;n \,,
\label{EinsteinAction4D}
\\ 
&&
q=\frac{n}{\sqrt{-g}}\,.
\label{q}
\end{eqnarray}
Here $n$ is the 4D analog of the particle density in the quantum vacuum (density of the “spacetime atoms”), which has the same dimension 4 as the tetrad determinant
\begin{equation}
[n] =[\sqrt{-g}]  =\frac{1}{[L]^4} \,,
\label{nDim}
\end{equation}
$q$ is the vacuum variable, and $\mu$ plays the role of the chemical potential in the vacuum thermodynamics.

For illustration, we consider the metric of the expanding Friedmann-Robertson-Walker universe:
\begin{eqnarray}
ds^2=g_{\mu\nu}dx^\mu dx^\nu= - d\tau^2 + a^2(\tau) d{\bf r}^2\,,
\label{metric}
\\
H(\tau)=\frac{da/d\tau}{a(\tau)}\,,
\label{Hubble}
\end{eqnarray}
where $\tau$ is the conformal time; $a(\tau)$ is the scale factor; and $H(\tau)$ is time-dependent Hubble parameter.

The scale factor $a(\tau)$ has dimensions 1:
\begin{equation}
 [a(\tau)] =\frac{1}{[L]} \,,
\label{aDim}
\end{equation}
while the following quantities have zero dimension ($1/[L]^0=[1]=0$):
\begin{eqnarray}
 [q]=[\mu] =[\epsilon]=[R]=[G_N]=[\Lambda_0]=
 \nonumber
 \\
 =[H]=[\tau]=[\psi]=[M]=[1]=0 \,.
\label{Odim}
\end{eqnarray}

Some of the dimensionless quantities can be fundamental, or correspond to some integer valued topological invariants.
For example, the "chemical potential"  $\mu$ may correspond to the topological invariant, $\mu=\pm 1$, which changes sign at the Big-Bang quantum phase transition.\cite{KlinkhamerVolovik2022} Since masses of particles are dimensionless,
and there is no fundamental mass scale, one can choose any convenient mass as a unit mass.

Note also that the dimensionless interval in Eq.(\ref{DimensionInterval}) does not mean the existence of the fundamental length, such as Planck length. First, because the gravitational coupling  $1/G_N$ is not fundamental (see, however, Sec.\ref{BINY}). Second, in the model of the superplastic (malleable) vacuum there is no equilibrium value of the distance between the neighbouring lattice points.  As distinct from the solid state crystals, arbitrary deformations of the vacuum crystal are possible.
Also, in the Diakonov model,\cite{Diakonov2011} the metric is the emergent phenomenon, while on the fundamental level the distance between the spacetime points is not determined.

\subsection{Black hole entropy}

In terms of the dimensionful metric, the acceleration is dimensionless:\cite{Volovik2021} 
\begin{equation}
a^2  = g_{\mu\nu} \frac{d^2x^\mu}{ds^2} \frac{d^2x^\nu}{ds^2} \,,
\label{acceleration1}
\end{equation}
\begin{equation}
 [a^2] = [g_{\mu\nu}] [x^\mu] [x^\nu]= \frac{1}{[l]^2} \cdot [l]^2=[1] =0\,.
\label{acceleration2}
\end{equation}
This leads to the dimensionless Unruh temperature:
\begin{equation}
T_{\rm U}=\frac{a}{2\pi}\,\,, \,\, [T_{\rm U}] = [1] =0\,.
\label{Unruh}
\end{equation}
The same is valid for the Gibbons-Hawking temperature of the cosmological horizon. As follows from Eq.(\ref{Odim}), it is also dimensionless:
\begin{equation}
T_H=\frac{H}{2\pi}\,\,, \,\, [T_H]=[H] = [1] =0\,,
\label{GibbonsHawking}
\end{equation}
Both equations, (\ref{Unruh}) and (\ref{GibbonsHawking}),  look fundamental: they do not contain parameters.

However, for the temperature of the Hawking radiation from the black hole horizon, $T_{\rm BH}=1/8\pi G_N M$, situation is different.  Although the Hawking temperature is dimensionless ($[T_{\rm BH}] = [1]$, since $[G_N]=[M]=[1]$), it is not fundamental, because it depends on the parameter $G_N$, which is not fundamental (see, however, Sec. \ref{BINY}). The same concerns the conventional Bekenstein-Hawking entropy, 
\begin{equation}
S_{\rm BH}= \frac{A}{4G_N}   \,.
\label{EntropyConventional}
\end{equation}
The entropy in Eq.(\ref{EntropyConventional}) is dimensionless, since both $G_N$ and the area $A$ are dimensionless.\cite{Volovik2021}   
That the horizon area is dimensionless can be seen from the general equation for area:
\begin{equation}
dA=\sqrt{dS^{ik} dS_{ik}} \,\,,\,\,[A]=[1]=0 \,,
\label{Area1}
\end{equation}
and from the simple spherical case:
\begin{equation}
A=\int_0^\pi d\theta \int_0^{2\pi} d\phi \,\sqrt{g_{\phi\phi} g_{\theta\theta}}\,\,,\,\, [g_{\phi\phi}]=[g_{\theta\theta}]=[A]=0 \,.
\label{Area2}
\end{equation}

The Bekenstein-Hawking entropy (\ref{EntropyConventional}) determines the black hole thermodynamics, but similar to the Hawking temperature  it does not look as fundamental, since it contains the gravitational coupling $1/G_N$.  Also it is not clear why the microscopic degrees of freedom responsible for the black hole entropy should be characterized by the Planck length.\cite{Verlinde2022} In the superplastic vacuum approach,\cite{KlinkhamerVolovik2019} the Planck length scale is simply absent, since there is no equilibrium value of the distance between the lattice points: the superplastic vacuum can be arbitrarily deformed. 

On the other hand, since the area is dimensionless, one may suggest that the entropy of the black hole horizon can be expressed in terms of the area only:
\begin{equation}
S_{\rm BH}= \eta A \,\,,\,\,   [\eta]=[S_{\rm BH}]=[A]=[1]=0 \,.
\label{EntropyArea}
\end{equation}
Here $\eta$ is some fundamental dimensionless parameter, like the topological invariant. 
In this case one may take the point of view that Einstein’s gravity equations can be derived solely from thermodynamics.\cite{Jacobson1995} 
 Then the constant of proportionality $\eta$  between the entropy and the area determines the gravitational coupling, $1/G_N= 4\eta$.
In this thermodynamic approach, $1/G_N$ may become fundamental due to the fundamentality of the parameter $\eta$.

However, in the thermodynamic approach to gravity there is the “species problem”,\cite{Jacobson1994} i.e.  the gravitational coupling $G_N$ may depend on the number of fermionic and bosonic fluctuating quantum fields.\cite{Sakharov1968,FFZ1997,Visser2002} In principle, with some relations between the numbers of quantum fields, $1/G_N$ can be negative, or even zero. This destroys many conjectures, which are based on positivity of the gravitational coupling.\cite{Volovik2022}  If so, $1/G_N$ cannot be the fundamental parameter.

In principle, this “no-go theorem" can be avoided, if  $1/G_N$ is  the quantum number, which is related to symmetry and/or topology. 
In this case the parameter $1/G_N$ does not depend on the interaction between gravity and quantum field, though it may experience jumps during the topological quantum phase transitions. The latter takes place in topological materials when one varies the parameters of interaction,\cite{Volovik2007,Volovik2018} and may take place when the Universe crosses the Big Bang.\cite{KlinkhamerVolovik2022}

\subsection{Einstein-Cartan, Barbero-Immirzi, Nieh-Yan and topology}
\label{BINY}

The topological invariants, which are  relevant for the quantum vacuum,  are known  in the crystalline matter,\cite{NissinenVolovik2019,Nissinen2020,NissinenVolovik2021}  and thus they can be extended to the superplastic  quantum vacuum. The topology in the crystalline quantum vacua is enriched due to the dimensional elasticity tetrads in Eq.(\ref{ElasticityTetrads}), which come from the geometric $U(1)$ phases. This topological apptoach is not applicable to the conventional Einstein gravity, where tetrads are absent. But it may take place in the Einstein-Cartan gravity (Einstein–Cartan–Sciama–Kibble theory), which is expressed in terms of tetrads, and thus is more fundamental than the conventional Einstein gravity, which is based on metric. Such type of gravity emerging in superplastic crystals has been discussed in Ref. 
\cite{Zubkov2019}.

The gravitational action in the Einstein-Cartan gravity can be expressed in terms of the differential forms, which contain the elasticity tetrads as the translational gauge fields:
\begin{equation}
S_{\rm EC} \sim \epsilon_{abcd}\int d^4 x\,E^a \wedge E^b\wedge R^{cd} \,.
\label{GravityAction}  
\end{equation}
 This action is dimensionless because the one-form tetrad has dimension 1, $[E^a_\mu]=\frac{1}{[L]}$, while the curvature two-form $R^{ab}$ has dimension 2:
\begin{equation}
 [R^{ab}_{\mu\nu}]=\frac{1}{[L]^2}\,.
\label{GravityDimension}  
\end{equation}
That is why with the dimensional elasticity tetrads,  the topology of the $3+1$ the crystalline phases,\cite{NissinenVolovik2019,Nissinen2020,NissinenVolovik2021} may provide the fundamental topological prefactor in Eq.(\ref{GravityAction}), with $1/G_N$ being integer or fractional topological number.

The same can be valid for the dimensionless parameter in the Barbero-Immirzi action:
\begin{equation}
S_{\rm BI} \sim \int d^4 x\,E^a \wedge E^b\wedge R_{ab} \,.
\label{BarberoImmirzi}  
\end{equation}
Eq.(\ref{BarberoImmirzi}) looks similar to the Nieh-Yan term in the action, see e.g. Ref. \cite{Calcagni2009}. Due to dimensional tetrads, $[E^a_\mu]=\frac{1}{[L]}$, the prefactors in the Nieh-Yan and in the Barbero-Immirzi actions are dimensionless, and thus can be fundamental.\cite{Volovik2021} It is not excluded that these parameters are the topological invariants similar to that in topological insulators, semimetals and superconductors.\cite{NissinenVolovik2019} 

The dimensional metric and tetrads appear also in such topological field theories as the $BF$-theoryl. For example, the composite metric (Sch\"onberg-Urbantke metric \cite{Schonberg1971,Urbantke1984,Jacobson1991,Obukhov1996,HehlObukhov2003,Friedel2012})  can be formed by the triplet of the 2-form fields ${\bf B}^a$:
\begin{equation}
\sqrt{-g}g_{\mu\nu}=\frac{1}{12}  e_{abc}e ^{\alpha\beta\gamma\delta} B^a_{\mu\alpha} B^b_{\beta\gamma} B^c_{\delta \nu} \,.
\label{MetricB}  
\end{equation}
The 2-forms in the $BF$ action $\int\, B \wedge F$ have dimension 2,  $[B]=[F]=1/[L]^2$. That is why the composite metric in Eq.(\ref{MetricB}) has also dimension 2, $[g_{\mu\nu}]=1/[L]^2$.
In the same way, the two-form field $B$ can be represented as the bilinear combination of the tetrads,\cite{Jacobson1991} $B=E\wedge E$. These one-form tetrads have dimension 1, $[E^a_\mu]=1/[L]$.

\subsection{ADM formalism and fine structure constant}
\label{ADMsec}

The  Arnowitt-Deser-Misner (ADM)  formalism\cite{ADM2008} is used for the  Hamiltonian formulation of general relativity. Let us consider this formalism and its application using  the dimensional metric. One has the following metric elements and their dimensions:
 \begin{eqnarray}
g_{ik}=\gamma_{ik}\,\,,\,\,  [\gamma_{ik}] =\frac{1}{[L]^2}\,,
\label{g_ik}
\\
g_{0i}=N_i= \gamma_{ik}N^k \,\,,\,\,  [N_i] =\frac{1}{[L]^2} \,\,,\,\,  [N^i] =0\,,
\label{g_0i}
\\
g_{00}= \gamma_{ik}N^iN^k - N^2 = N^i N_i - N^2  \,\,,\,\,  [N] =\frac{1}{[L]} \,,
\label{g_00}
\\
g^{00}= - \frac{1}{N^2} \,\,,\,\,  [g^{00}] =[L]^2\,,
\label{g^00}
\\
g^{0i}=  \frac{N^i}{N^2}\,\,,\,\,  [g^{0i}] =[L]^2\,,
\label{g^0i}
\\
g^{ik}= \gamma^{ik}- \frac{N^iN^j}{N^2}\,\,,\,\,  [\gamma^{ik}] =[L]^2\,,
\label{g^ik}
\\
\sqrt{-g}=N\sqrt{\gamma}\,\,,\,\,  [\sqrt{\gamma}\,] =\frac{1}{[L]^3}\,,
\label{g}
\\
\gamma^{ik} \gamma_{kl}= \delta^i_l\,.
\label{gammagamma}
\end{eqnarray} 
Here $N$ and $N^i$ are lapse and shift functions correspondingly, and $\gamma_{ik}$ are the space components of the metric.
Actually the ADM metric is similar to the acoustic metric in Sec.\ref{SecAcoustic}, where the role of the shift vector $N^i$ is played by the velocity $v^i$ of the liquid.

The ADM formalism  allows to consider dynamics in curved space in terms of the Poisson brackets.
Let us consider this on example of  Poisson brackets for the classical $3+1$ electrodynamics in curved space:
\begin{eqnarray}
\{   A_i({\bf r}), D^k({\bf r'})\}  =\delta^k_i \delta({\bf r}-{\bf r}')\,,
\label{PBEM3}
\end{eqnarray} 
which in terms of the gauge invariant fields is:
\begin{eqnarray}
\{   B^i({\bf r}), D^k({\bf r'})\}  =e^{ikl}\nabla_l \delta({\bf r}-{\bf r}')\,.
\label{PBEM4}
\end{eqnarray} 
Here ${\bf B}$ is magnetic field, and the vector ${\bf D}$ is the electric induction of the quantum vacuum  (electric displacement field). 
The electric induction ${\bf D}$ is expressed in terms of the electric field $E_i=F_{0i}$:
\begin{equation}
D^k =\frac{1}{\alpha} \frac{\sqrt{\gamma}}{N}    \gamma^{ik} E_i \,.
\label{D}
\end{equation}
Here $\alpha$ is the dimensionless fine structure constant, which determines
the dielectric constant -- the electric permittivity of the relativistic quantum vacuum, $\epsilon_{\rm vac}$, and the magnetic permeability of the vacuum, $\mu_{\rm vac}$:
\begin{equation}
\epsilon_{\rm vac}=\frac{1}{\mu_{\rm vac}}=\frac{1}{\alpha}
\,.
\label{dielectric_constant}
\end{equation} 
Here we do not include the artificial factor $4\pi$. 

In spite of the dimensional metric, the electric inductance ${\bf D}$ has the same dimension 2 as the electric field ${\bf E}$:
\begin{equation}
 [D^i] =[E_i]=\frac{1}{[L]^2} \,.
\label{Ddimension}
\end{equation}
This follows from Eqs.(\ref{g_ik}), (\ref{g^00}) and (\ref{g}) for dimensions of the ADM metric elements in the $3+1$ spacetime.

The corresponding quadratic Hamiltonian for the electromagnetic field is:
\begin{equation}
H = \int  \frac{d^3r}{2}\frac{N}{\sqrt{\gamma} }\gamma_{ik} \left( \alpha D^i D^k + \frac{1}{\alpha} B^i B^k\right)
\,.
\label{HEB}
\end{equation} 
The Hamiltonian has dimension 1, i.e. $[H]=1/[L]$.
Both the Hamiltonian in Eq.(\ref{HEB}) and the Poisson bracket in Eq. (\ref{PBEM4}) do not contain the gauge potentials. The 
gauge potentials also do not enter the Poisson brackets for charged particle, $\{  p_i,p_j \} = q F_{ij}$ and $\{p_i({\bf r}), D^k({\bf r'})\}  =-q\delta^k_i \delta({\bf r}-{\bf r}')$, where $q$ is the dimensionless electric charge of the particle in terms of the electric charge of the electron.

The quantization of electromagnetic field is obtained by the substitution of the Poisson brackets (\ref{PBEM4}) by commutation relations between ${\bf D}$ and ${\bf B}$.

The Poisson brackets in Eqs. (\ref{PBEM3}) and (\ref{PBEM4}) look as fundamental. They do not depend on the metric and  do not contain physical  parameters of the quantum vacuum. However, the function ${\bf D}$ in Eq.(\ref{D}) breaks this fundamentality.  It is the phenomenological variable, which describes the response of the vacuum to electric field. This response contains the electromagnetic coupling $1/\alpha$, which is not fundamental because of the corresponding  “species problem”: it depends on the fluctuating bosonic and fermionic fields in the quantum vacuum, and is space-dependent. While according to Sec. \ref{BINY}, the gravitational coupling $1/G_N$ can be fundamental due to topology, there are no topological invariants which could support the fundamentality of the electromagnetic coupling $1/\alpha$.
This is in favour of the scenario in which the quantum electrodynamics is the effective low-energy theory, where for example the gauge fields emerge as the bilinear combinations of the fermionic fields, or/and the gauge fields emerge in the vicinity of the topologically stable Weyl points in the fermionic spectrum.\cite{Volovik1986a,Volovik1986b,Volovik2003,Horava2005} 

This, however, does not exclude the other possible pre-quantum and pre-spacetime theories, see Ref. \cite{Singh2022} and references therein.

\section{Conclusion}

Several approaches to quantum gravity (including the model of superplastic vacuum; Diakonov tetrads emerging  as the bilinear combinations  of the fermionis fields; $BF$-theories of gravity; and effective acoustic metric)  suggest that in general relativity  the metric has dimension 2, i.e. $[g_{\mu\nu}]=1/[L]^2$, irrespective of the dimension of spacetime. One consequence of such dimension of the metric is that the wave function in quantum mechanics is dimensionless, $[\psi(x)]=[1]=0$. This also leads to the dimensionless quantum fields.

On the other hand, if one starts with the conjecture that in quantum mechanics the wave function is naturally dimensionless,  one obtains that the metric acquires the dimension 2. This suggests the close connection between quantum mechanics and general relativity.

 {\bf Acknowledgements}.  This work has been supported by the European Research Council (ERC) under the European Union's Horizon 2020 research and innovation programme (Grant Agreement No. 694248).

\end{document}